\documentclass[pdflatex,sn-mathphys-num]{sn-jnl}

\usepackage{graphicx}%
\usepackage{multirow}%
\usepackage{amsmath,amssymb,amsfonts}%
\usepackage{amsthm}%
\usepackage{mathrsfs}%
\usepackage[title]{appendix}%
\usepackage{xcolor}%
\usepackage{textcomp}%
\usepackage{manyfoot}%
\usepackage{booktabs}%
\usepackage{algorithm}%
\usepackage{algorithmicx}%
\usepackage{algpseudocode}%
\usepackage{listings}%

\usepackage{bm}
\usepackage[mathlines]{lineno}
\usepackage{upgreek}
\usepackage{longtable}
\usepackage{xcolor}
\usepackage{gensymb}

\usepackage[export]{adjustbox}
\usepackage{wrapfig}
\usepackage{amsmath}
\usepackage{multirow}
\usepackage{tabularx}
\usepackage{gensymb}
\usepackage{mathtools}
\usepackage{siunitx}

\usepackage{wasysym}
\usepackage{color}
\usepackage{float}
\usepackage{comment}

\usepackage{ragged2e}
\usepackage{color,soul}

\usepackage{comment}
\usepackage{url}

\usepackage[mathlines]{lineno}

\theoremstyle{thmstyleone}%
\theoremstyle{thmstyletwo}%

\theoremstyle{thmstylethree}%

\begin{document}


\title[Article Title]{Crowding and confinement of mesoscale swimmers in living droplets}







\author[1,2]{\fnm{L.} \sur{Malik}}
\equalcont{These authors contributed equally to this work.}

\author[1]{\fnm{N.} \sur{Sharadhi}}
\equalcont{These authors contributed equally to this work.}

\author[1]{C. P. Llarena}
\author[1]{M. Lamminmäki}
\author[1]{R. A. Lara}
\author[3]{V. Jokinen}
\author[4]{M. Lisicki}
\author*[1]{M. Backholm}

\affil[1]{\orgdiv{Department of Applied Physics}, \orgname{Aalto University}, \orgaddress{\city{Espoo}, \country{Finland}}}

\affil[2]{\orgdiv{Department of Mechanical Engineering}, \orgname{IIT Gandhinagar}, \orgaddress{\country{India}}}

\affil[3]{\orgdiv{Department of Chemistry and Materials Science}, \orgname{Aalto University}, \orgaddress{\city{Espoo}, \country{Finland}}}

\affil[4]{\orgdiv{Faculty of Physics}, \orgname{University of Warsaw}, \orgaddress{\city{Warsaw}, \country{Poland}}}

\email{matilda.backholm@aalto.fi}


\abstract{

Droplets encapsulating active or living matter provide controllable platforms for studying collective dynamics and active interfacial actuation. However, prior work has largely focused on quasi-2D confinement, microscopic swimmers, or externally driven, non-living droplets. Here, we introduce living droplets internally activated by tens of enclosed mesoscale swimmers to study crowding and confinement in 3D with swimmer sizes comparable to the droplet radius. In large droplets with few swimmers, the droplet oscillation frequencies collapse onto the classical sessile-drop frequency-volume scaling. By increasing the number of swimmers or decreasing the droplet size, crowding and confinement alter the droplet oscillation frequencies and amplitudes as many-body or swimmer-droplet interactions start dominating. We develop a scaling-law framework linking the oscillation amplitude to swimmer speed through collision probability and momentum transfer, enabling indirect 3D velocimetry in the opaque, crowded droplets. Complementary quasi-2D experiments with direct tracking reveal dimensionality-dependent crowding effects, which we successfully describe with a kinetic theory-based model. We show distinct crossovers and strong-crowding asymptotic scalings of the swimmer speed with the dimensionless swimmer density, $U_{\mathrm{crowd}}\sim(N^{\ast})^{-1}$ in 3D and $U_{\mathrm{crowd}}\sim(N^{\ast})^{-1/2}$ in quasi-2D. Our results connect mesoswimmer encounter dynamics to macroscopic droplet mechanics and inform the design of bio-inspired micromachines, fluidic control, and sensing technologies. } 

\maketitle

\section{Introduction}\label{sec1}
Droplets enclosing out-of-equilibrium living or artificial swimmers~\cite{Elgeti2015,Bechinger2016} present a compelling system for exploring collective dynamics in active, living, and soft matter physics~\cite{Klotsa2019,Marchetti2013}. Previous studies have focused on confined and/or crowded non-living active matter, such as active nanoparticles~\cite{Kokot2022}, motile colloids~\cite{Bricard2013,Vutukuri2020}, active network of microtubules~\cite{Sciortino2025}, acousto-magnetic microswimmers~\cite{Ahmed2017}, robots~\cite{Boudet2021} and robotic fish~\cite{Boudet2025}, as well as living organisms, such as fish~\cite{Ventejou2024,Lafoux2024}, bacteria~\cite{Wioland2013, Lushi2014,Soto2014,Villalobos2025}, algae~\cite{Ostapenko2018,Cammann2021,Mondal2021,Tainio2021,Wei2024}, spermatozoa~\cite{Kantsler2013}, and theoretical microswimmers~\cite{Soto2014,Sprenger2020,Kawakami2025,Kawakami2025b}. These have revealed how concerted motion of geometrically constrained internal swimmers can drive droplet shape fluctuations and self-propulsion, as well as organized fluid flows. Most studies have focused on the effects of crowding and confinement in quasi-2D using cylindrical disks or planar surfaces. Recently, the 3D confinement of microscopic swimmers has gained attention, with bacteria confined in droplets~\cite{Villalobos2025} and 3D chambers~\cite{Wei2024} as well as active particles~\cite{Vutukuri2020} and microtubules~\cite{Sciortino2025} deforming giant lipid vesicles. Studying the effect of 3D confinement is of great fundamental interest since most organisms naturally swim in 3D. Here, we combine experiments with theoretical modelling to understand the effects of crowding in 3D droplets (radius $R$) with mesoscale swimmers of lengths $L\sim R$. This marks a significant departure from prior research predominantly focused on 2D confinement and/or much smaller swimmer sizes relative to their confinement ($L\ll R$).

Pearl-like liquid droplets on superhydrophobic substrates~\cite{Quere2005} represent excellent spherical experimental microcompartments for probing 3D confinement. The mechanical response of sessile droplets on such surfaces has been described with classical resonance models, which relate the fundamental mode oscillation frequency to droplet volume, contact angle, and surface tension~\cite{Noblin2004,McHale2005,Celestini2006,Sharp2011,Bostwick2014,Chang2015,Sakakeeny2021}. Yet, they have exclusively focused on externally activated, non-living droplets, and there is no understanding of how the classical models generalize to droplets that are internally activated by living swimmers. We expect measurable effects, especially in the mesoscale regime where the size of the swimmer $L$ is comparable to the droplet radius $R$. This system is not only relevant to living and soft matter physics, but also supports the upscaling of droplet-based sensors~\cite{Kaminski2016} and swimmer-driven micro-machines~\cite{Hiratsuka2006,Leonardo2010,Oda2024} to the mesoscale. Furthermore, a wide range of living micro- to meso-organisms live as aeroplankton within atmospheric water droplets~\cite{Smith2013}, affecting the weather~\cite{Christner2008} and public health through, for example, allergies and as transmission vectors~\cite{Smith2013}. How actively swimming meso-organisms influence the mechanics of such droplets remains unstudied.

\begin{figure}[!t]
\center
{\includegraphics[width=1\linewidth]{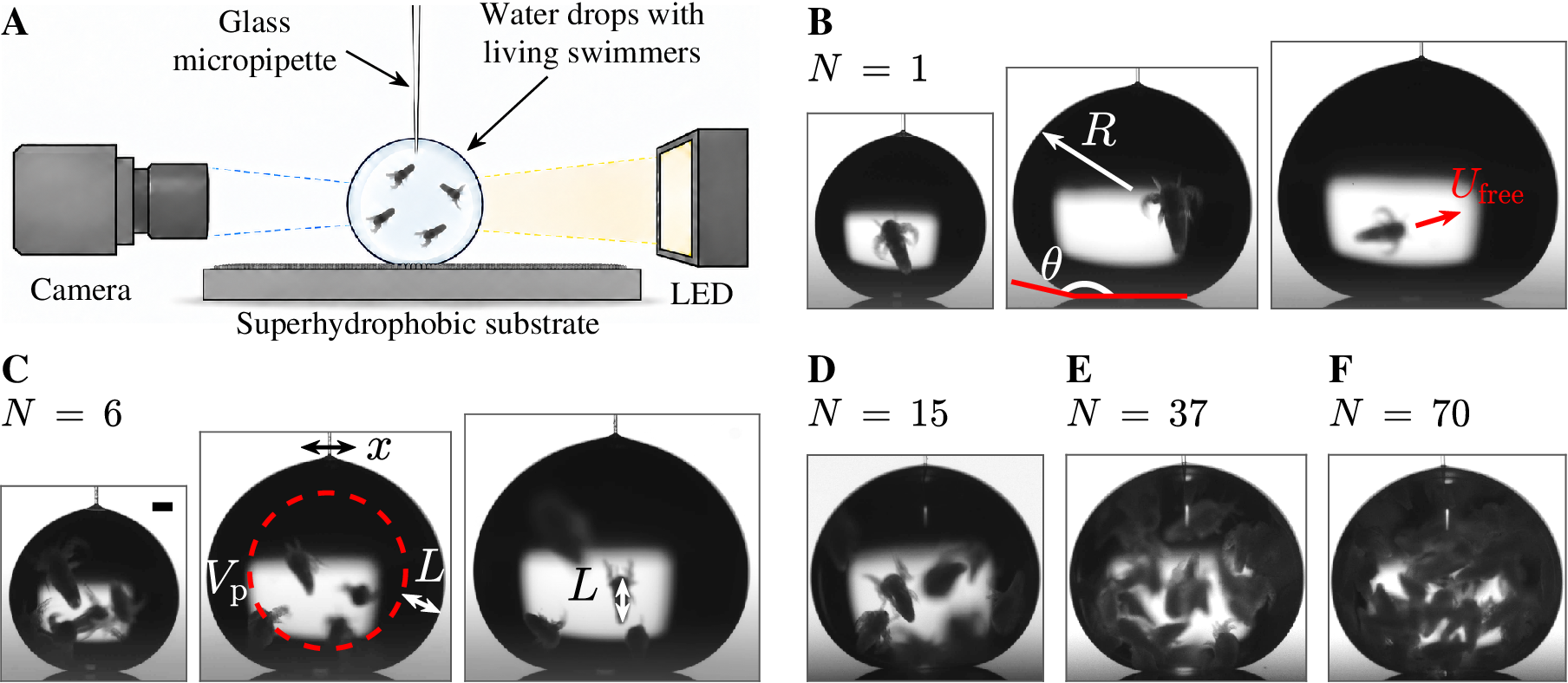}}
\caption
{\textbf{Experiments in 3D living droplets.}
(A) Schematic (not to scale) of 3D living droplets experiments. (B-F) Experimental images of living droplets supported by a superhydrophobic substrate (contact angle $\theta=163^\circ\pm3^\circ$) with (B) $N=1$, (C) $N=6$, and (D-F) $N=15-70$ \textit{Artemia} swimmers of length $L=490\pm20\ \mu$m. Images are shown for droplet volumes $V= 3-13.5\ \mu$L, corresponding to radii $R= 895-1550\ \mu$m and $L/R= 0.30-0.57$. The vertical rod at the top of the droplet is a glass micropipette; the droplet oscillations are measured from its deflection, $x$. The peripheral volume $V_{\mathrm{p}}$ indicates where swimmer-drop collisions can happen. Scale bar $200 \ \mu$m.
}
\label{fig:fig1}
\end{figure}

In this work, we explore a new class of living droplets where the motion of mesoscale swimmers creates macroscopic oscillations of round sessile droplets on a superhydrophobic black silicon surface (Fig.~\ref{fig:fig1} and Supplementary Movies S1$-$S9). We focus at $L\sim R$ in 3D, where the effects of momentum transfer, collision dynamics, and confinement will be very different than at the previously studied $L\ll R$ regime in quasi-2D. Importantly, our swimmers are large enough to cause small changes in the drop position upon a collision. Due to the major optical challenges associated with observing individual organisms swimming inside a 3D droplet, we develop a new experimental approach to probe the intricate many-body system through the active oscillations of the living droplets. We also perform swimming experiments in the classical quasi-2D droplet geometry to compare our 3D experiments with. We primarily use \textit{Artemia} nauplii as a model organism for their well-characterized butterfly swimming mechanism~\cite{Azra2022,Williams1994,backholm_artemia}, and complement our study with adult copepods for their distinct intermittent swimming strategies characterized by powerful swimming bursts~\cite{Svetlichny2020} (see Supplementary Note 1: Supplementary Fig. S1). At increased swimmer confinement and crowding, we observe deviations from a classical resonance model~\cite{Sakakeeny2021,Celestini2006}, originally developed for non-living droplets. We derive a scaling law model to determine the average swimming speed of the meso-organisms based on the average droplet oscillation amplitudes. We report a clear difference between how crowding affects the swimming velocity of \textit{Artemia} in 3D compared to quasi-2D geometries, which we explain using a kinetic-theory-inspired model.

\section{Results and discussion}\label{sec2}

\subsection*{Living droplet experiments}

The experiments are briefly described below, see \textit{Methods} for all technical details and Supplementary Note 2: Data analysis. Droplets of brine and mesoswimmers are placed on a superhydrophobic black silicon substrate coated with fluoropolymer (Fig.~\ref{fig:fig1}A)~\cite{Sainiemi2011}. A range of different droplets volumes $V=2-20\ \mu$L (radii $R= 600-1750\ \mu$m) and number of swimmers ($N=1-70$) was studied. We focus mostly on $24$h old \textit{Artemia} nauplii swimmers with lengths of $L=490\pm20\ \mu$m, but have also performed experiments on $48$h \textit{Artemia} ($L=710\pm30\ \mu$m) as well as several drops with an adult \textit{Acartia tonsa} copepod ($L=1120\pm20$ $\mu$m; Supplementary Fig. S1). The nearly spherical droplets have an average contact angle of $\theta=163^\circ\pm3^\circ$ and extremely low friction force~\cite{Backholm2020,Backholm2024}. A vertical glass micropipette is used to assist in keeping the droplet in place and in detecting interface oscillations (Supplementary Fig. S2--S4). For each swimmer type, micropipettes with similar spring constants are used ($5.2\pm0.1$ nN/$\mu$m for \textit{Artemia} and $2.44\pm0.03$ nN/$\mu$m for copepods; see \textit{Methods}, Supplementary Fig. S5, and Ref.~\cite{Backholm2019} for details on the calibration), ensuring that the micropipettes do not influence the droplet oscillations (Supplementary Movie S10 and Fig. S6). The experiments are observed at 40--100 fps at 2x magnification with a side-view camera using back-illumination with an LED. The entire setup (Supplementary Fig. S7) is mounted on a vibration-isolated table and shielded with a black hardboard box. Image analysis quantifies droplet geometry and the micropipette deflection ($x$) as a function of time. Representative micrographs for $N=1$, $N=6$, and $N=15-70$ \textit{Artemia} ($24$h old) are shown in Fig.~\ref{fig:fig1} (Supplementary Movies S1$-$S9, S11; see Supplementary Fig. S8 and Movie S12 for examples with 48h swimmers), with $L$, $R$, $\theta$, and micropipette deflection $x$ indicated.

\begin{figure*}[!t]
\center{
\includegraphics[width=1\linewidth]{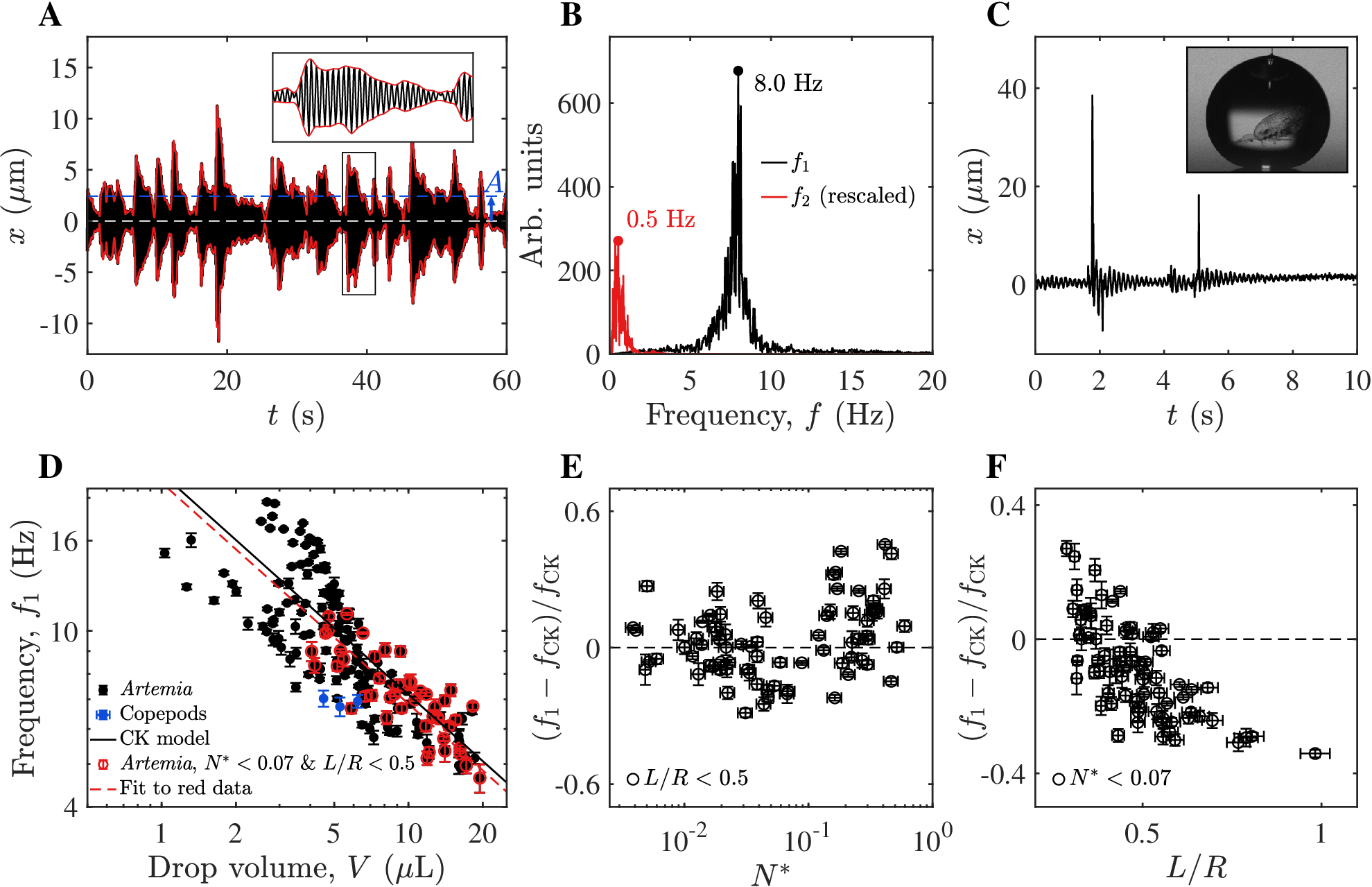}}
\caption
{\textbf{Living droplet oscillations.}
(A) Deflection versus time for a typical experimental case of $N=6$ \textit{Artemia} swimmers ($48$h) inside a droplet of $\ V\approx 4 \ \mu$L. The amplitude $A$ (rms of $x$) is shown with a dashed blue line. Inset shows a zoomed-in view of an oscillation envelope. 
(B) Frequency spectra corresponding to the droplet deflection and the oscillation envelope, showing the dominant frequencies $f_1$ and $f_2$, respectively. The $f_2$ spectrum is rescaled for visualization. 
(C) Deflection versus time for a droplet containing a single copepod. Inset shows an image from a copepod experiment ($R = 1.16 \text{ mm}$).
(D) $f_{\mathrm{1}}$ plotted as a function of droplet volume for varying $N$ and age of \textit{Artemia} (black markers) and for $N=1$ copepod droplets (blue markers). The uncrowded and unconfined \textit{Artemia} droplets are plotted in red and the experimental fit to these points gives $f_{\mathrm{1}}=(21.6\pm0.9)\cdot V^{-0.5}$ (dashed red line, error $95\%$ confidence of fit), in good agreement with the CK model $f_{\mathrm{CK}}=(22.7\pm0.3)\cdot V^{-0.5}$~\cite{Celestini2006,Sakakeeny2021} (black line). The measured $f_1$ deviates differently from $f_{\mathrm{CK}}$ as a function of (E) increased crowding and (F) confinement. The errors are standard deviations and propagated errors based on these.
}
\label{fig:fig2}
\end{figure*}

The detection of droplet activity using the micropipette leads to a typical oscillatory signal as shown in Fig.~\ref{fig:fig2}A (Supplementary Movie S13), where the \textit{Artemia} drive continuous drop oscillations. The droplet amplitude, $A$, is defined as the root-mean-square (rms) of $x$. The noise on a micropipette in a water droplet without swimmers, was in the range of $x\approx \pm 0.3$ $\mu$m (Supplementary Fig. S9), and only datasets with amplitudes exceeding this ($A>0.4$ $\mu$m) were included. A Fourier transform on the $x(t)$ data quantifies the droplet oscillation frequencies (Fig.~\ref{fig:fig2}B, and Supplementary Figs. S10--S12). For \textit{Artemia}, we observe two distinct peaks: $f_{\mathrm{1}}$, associated with the fundamental mode of resonance and $f_{\mathrm{2}}$, associated with the formation of envelopes (highlighted in red in Fig. \ref{fig:fig2}A). The lower frequency response, $f_{\mathrm{2}}$ may be attributed to the occurrence of simultaneous or synchronized impact events, at times from multiple swimmers, collectively enhancing the droplet oscillatory motion. We find $f_{\mathrm{1}}$ to be roughly one order of magnitude higher than $f_{\mathrm{2}}$, but observe no connection between $f_{\mathrm{2}}$ and the level of confinement or crowding (see Supplementary Fig. S13). We do not investigate this mode of oscillation in more detail. 

The swimming bursts of a copepod in a drop shows large peaks separated by damped harmonic oscillations (Fig.~\ref{fig:fig2}C; Supplementary Fig. S14 and Movie S14), which is a result of the burst-like swimming strokes followed by long pauses of adult copepods~\cite{Svetlichny2020}. Encapsulated microscopic \textit{Paramecia}, however, exhibited no visible droplet oscillations (Supplementary Movie S15), likely because their steady ciliary beating produces smooth propulsion without creating the large momentum transfer required for droplet oscillations. The measurable activity of the living droplets of this work is thus expected to occur with active (living or inanimate) swimmers with puller-type, back-and-forth swimming modes that are powerful enough to create a measurable momentum transfer to the droplet.
 
\subsection*{Droplet oscillation frequencies}

The measured $f_{\mathrm{1}}$ decreases as a function of $V$ for all experiments with varying $V$, $N$, and $L$ of \textit{Artemia} and few cases of copepods (Fig.~\ref{fig:fig2}D; see Supplementary Fig. S15 for better distinction between the different experiments). We define the level of swimmer confinement as $L/R$ and crowding as the dimensionless swimmer density $N^*=NV_{\mathrm{a}}/V$, where $V_{\mathrm{a}}$ is the swimmer volume. As shown in Fig.~\ref{fig:fig2}D, at low crowding and confinement, we find $f_{\mathrm{1}} =(21.6\pm0.9)\cdot V^{-0.5}$ ($V$ in units of $\mu$L) in good agreement with the Celestini and Kofman (CK) model ~\cite{Celestini2006,Sakakeeny2021} $f_{\mathrm{CK}} = \left[ {\gamma \, \eta\cdot \, (1 - \cos\theta)}/{(2 \pi \, \rho_d)} \right]^{1/2} \, V^{-0.5} =(22.7\pm 0.3)\cdot V^{-0.5}\) ($V$ in units of $\mu$L), where \( \eta = 0.52 \left( e^{0.99 (1 + \cos\theta)} - 1 \right) \),  $\rho_{\mathrm{d}}=1000$ kg m$^{-3}$ is the droplet density, $\gamma=0.072$ Nm$^{-1}$ the surface tension, and $\theta=163\pm3^\circ$ the contact angle in our work. This model assumes a Bond number $\mathrm{Bo}=\rho_{\mathrm{d}}\ gR^2/\gamma\approx 0$, which is similar to our experiments ($\mathrm{Bo} \sim 0.05-0.4 \ <1$). Adult copeopods and \textit{Artemia} nauplii cover similar ranges of swimming frequencies ($4-6$ Hz and $3-10$ Hz, respectively) but use very different swimming styles: copepods with a powerful beat–sink swimming mechanism and \textit{Artemia} with a continuous butterfly flapping of their antennae. Yet, this does not translate to different droplet oscillation frequencies. The different swimming kinematics as well as swimmer number and size should strongly influence the internal fluid flow in the droplets. In an extended model by Sakakeeny \textit{et al.}~\cite{Sakakeeny2021}, a kinetic energy term for the internal flow is included. However, for superhydrophobic surfaces and very low Bo, the difference between the Sakakeeny~\cite{Sakakeeny2021} and CK~\cite{Celestini2006} model is negligible and the effects of the internal flow are thus minimal. The collapse of the experimental data in the low $L/R$ and $N^*$ regime suggests that living droplets behave as externally driven non-living liquid droplets.
 
The deviation from the CK-model increases at higher crowding (increasing $N^*$; Fig.~\ref{fig:fig2}E, which only includes low-confinement data) and confinement (increasing $L/R$; Fig.~\ref{fig:fig2}F, which only includes low-crowding data). Interestingly, these two different physical constraints affect the measured $f_1$ in different directions. Crowding causes a slightly higher $f_1$ compared to the predicted natural resonant frequency $f_{\mathrm{CK}}$ at high $N^*$, likely caused by frequent unsynchronised collision events of the high number of swimmers, rendering a higher collision rate than the self-oscillation rate of the droplet. In contrast, an increased confinement renders a lower $f_1$ than $f_{\mathrm{CK}}$, which can be caused by the swimmer bodies being in almost constant contact with the droplet, damping its ability to self-oscillate.

\subsection*{Droplet oscillation amplitude}

To fully understand how the activity of the living droplets is affected by crowding and confinement, we turn to studying the effect of $L/R$ and $N^*$ on the droplet amplitude. We focus on modelling $A$ with a scaling-law approach, considering the impact probability and the collision mechanics of our swimmer-droplet system. The probability that an \textit{Artemia} bumps into the droplet interface is given by the ratio of the peripheral volume \(V_{\mathrm{p}}=V-(4\pi/3) (R - L)^3\) (Fig.~\ref{fig:fig1}C) to the total droplet volume, $P \sim V_{\mathrm{p}}/V$, assuming a spherical droplet for simplicity. This simplifies to \(P \sim 3 (L/R) - 3 (L/R)^2 + (L/R)^3\). Assuming momentum conservation during an impact between \textit{Artemia} and the droplet yields: \(\rho_{\mathrm{d}} V U_{\mathrm{d}} \sim \rho_{\mathrm{a}} V_{\mathrm{a}} U_{\mathrm{swim}}\), where \(\rho_{\mathrm{d}}\), \(\rho_{\mathrm{a}}\) and \(V\), \(V_{\mathrm{a}}\)  are the mass densities and volumes of the droplet and \textit{Artemia}, respectively, and \(U_{\mathrm{d}}\) is the droplet velocity right after impact and \(U_{\mathrm{swim}}\) the \textit{Artemia} velocity right before impact. We assume $\rho_{\mathrm{d}}\sim\rho_{\mathrm{a}}$. Considering \(f_{\mathrm{1}}\) to be the dominant frequency of oscillation, the droplet velocity can be approximated as \(U_{\mathrm{d}}\sim A_{\mathrm{0}} f_{\mathrm{1}}\) given that the interface would move a distance of \(\sim A_{\mathrm{0}}\) in a time span of \(\sim 1/f_{\mathrm{1}}\) upon a single-swimmer collision. This simplifies the momentum conservation equation to \(VA_{\mathrm{0}} f_{\mathrm{1}} \sim V_{\mathrm{a}}U_{\mathrm{swim}}\), which yields \(A_{\mathrm{0}} \sim U_{\mathrm{swim}}V_{\mathrm{a}} / (V f_{\mathrm{1}})\). We finally make the first-order assumption that the time-averaged droplet oscillation amplitude caused by many swimmers over a sufficiently long experiment scales with the drop amplitude caused by one swimmer, the probability of a droplet collision, as well as the number of swimmers in the droplet: \(A_\mathrm{t} \sim P A_{\mathrm{0}} N\), which gives
\begin{equation}
    A_\mathrm{t} \sim \dfrac{U_{\mathrm{swim}} V_{\mathrm{a}} N}{V f_{\mathrm{1}}} \left[ 3 \dfrac{L}{R} - 3 \left(\dfrac{L}{R}\right)^2 + \left(\dfrac{L}{R}\right)^3 \right].
    \label{eq:At}
\end{equation}

\begin{figure*}[!t]
\center{
\includegraphics[width=0.80\linewidth]{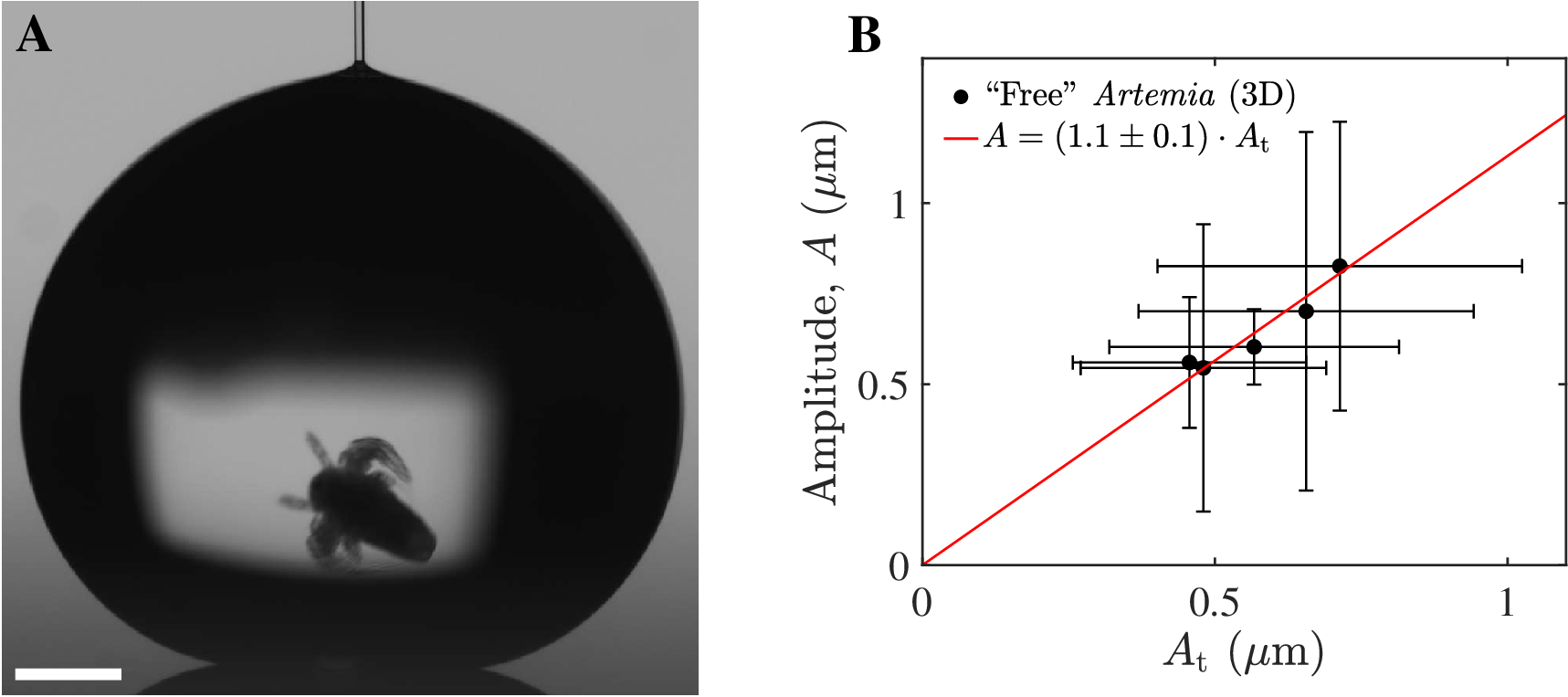}}
\caption
{\textbf{Droplet amplitude with \textit{Artemia} swimming ``freely".} (A) Image of ``free" \textit{Artemia} (24h) in droplet ($V\approx 7.6$ $\mu$L) with $L/R\approx0.39$ and $N^*\approx 0.006$. Scale bar 400 $\mu$m. (B) Comparison of average experimental amplitude, $A$ and theoretical amplitude, $A_\mathrm{t}$ (Eq.~\eqref{eq:At}) for the case of unconfined ($L/R<0.5$) and uncrowded ($N^*<0.01$) 24h \textit{Artemia} in 3D droplets, where we have assumed they are swimming at their free-swimming speed of $U_{\mathrm{free}}=1.2\pm0.5$ mm/s. The solid line is a linear fit of $A=(1.1\pm0.1)A_\mathrm{t}$, indicating very good agreement between the experimental data and theoretical model. The error bars are the experimental error of $A$ and the propagated error for $A_\mathrm{t}$.}
\label{fig:fig6}
\end{figure*}

We first investigate the case of unconfined ($L/R \lesssim 0.5$) and uncrowded ($N^*\lesssim 0.01$) 24h \textit{Artemia}, which we assume are swimming freely. We use swimming velocities measured in~\cite{backholm_artemia} on single \textit{Artemia} without confinement $U_{\mathrm{free}}=1.2\pm0.5$ mm/s and model the body as an ellipsoid $V_{\mathrm{a}}=\pi W L H/6$, with width $W=840\pm60$ $\mu$m and height $H=210\pm20$ $\mu$m (Supplementary Fig. S16). For the experiments (example in Fig.~\ref{fig:fig6}A) that obey the strict requirements of the swimmers being ``free'' (large enough $R$), yet able to cause drop oscillations above the noise level (small enough $R$), the ratio between the measured and calculated (Eq.~\eqref{eq:At}) amplitudes is $A_\mathrm{free}/A_\mathrm{t}(U_\mathrm{free})=1.1\pm0.1$ (Fig.~\ref{fig:fig6}B). We thus have quantitative agreement between our experiments and the model for the case of ``free'' swimmers where the swimming speed is known. This ratio corrects for the scaling-law nature of the model. 

\begin{figure*}[!t]
\center{
\includegraphics[width=0.75\linewidth]{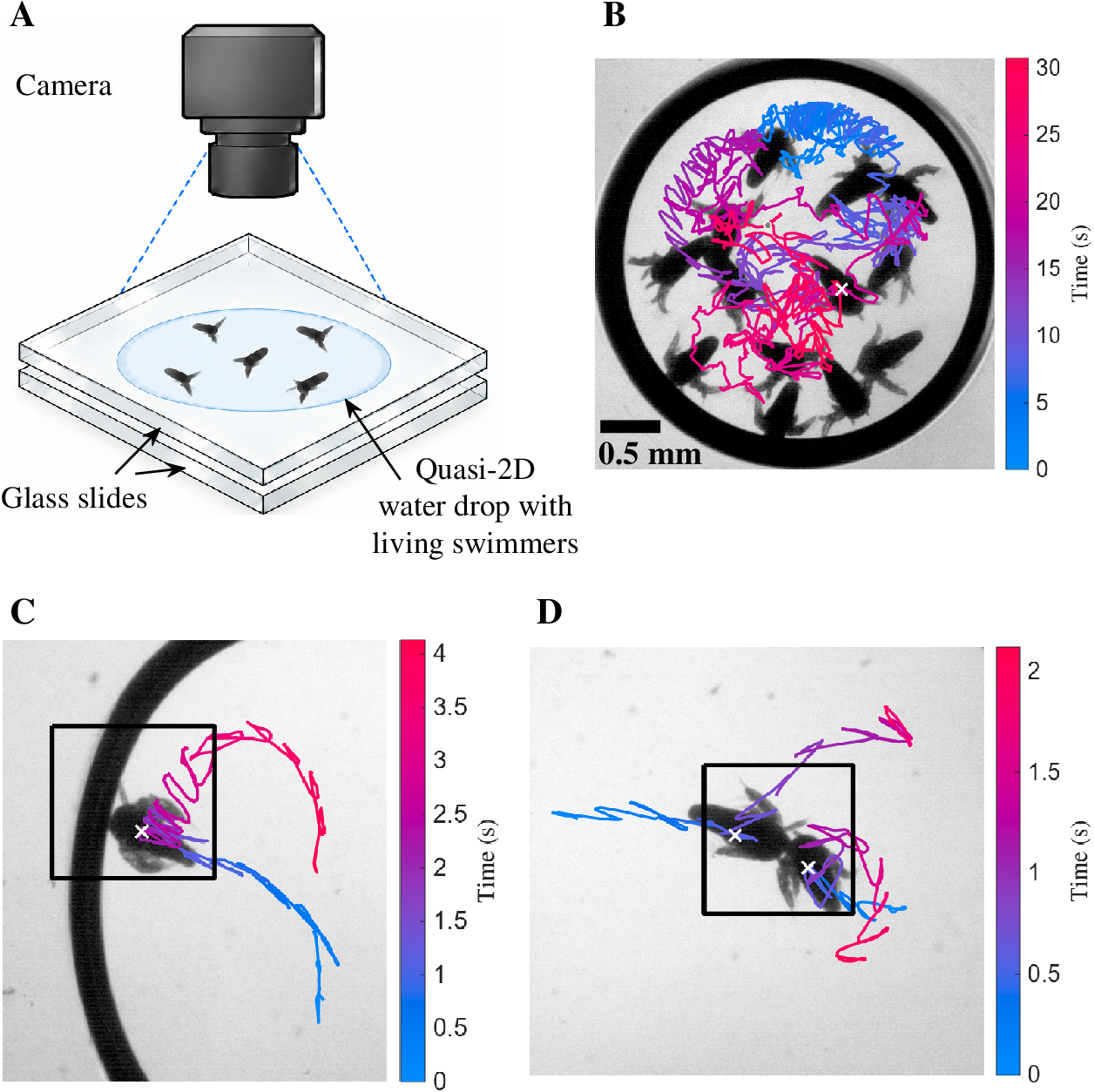}}
\caption
{\textbf{Quasi-2D swimming}
(A) Schematic (not to scale) of quasi-2D living droplet experiments with \textit{Artemia} (24h). (B) Example experiment with $N=13$ swimmers in an $R= 1.31\pm0.06$ mm confinement. The tracking of one swimmer is shown as a function of time. Representative experiments of an (C) \textit{Artemia}-wall collision (with $\tau_\mathrm{d}\approx 1.6$ s) and (D) \textit{Artemia}-\textit{Artemia} collision (with $\tau_\mathrm{a}\approx 0.8$ s). The blue-red colours show the \textit{Artemia} centroid position (white cross on the body in one time frame) over time and the black box indicates the collision area. The swimmers in (C-D) are $L\approx 490$ $\mu$m long for scale.
}
\label{fig:fig4}
\end{figure*}

\subsection*{Effects of confinement and crowding on swimming kinematics}
Crowding and confinement are known to reduce the speed of swimming~\cite{Boudet2025,Mondal2021}, with possibilities for complex steric interactions between living organisms~\cite{Backholm2014}. Due to the 3D geometry of our living droplets, we cannot track individual swimmers. Instead, we use the correlation between swimming speed and droplet oscillation amplitude of Eq.~\eqref{eq:At} to probe how \textit{Artemia} is affected by crowding or confinement: $U_{\mathrm{swim}}/U_{\mathrm{free}}=A/(1.1\cdot A_\mathrm{t}(U_\mathrm{free}))$, where the factor $1.1$ stems from the analysis of the ``free" swimmers, $A$ is the measured amplitude, and $A_\mathrm{t}(U_\mathrm{free})$ is the free-swimming benchmark amplitude calculated with Eq.~\eqref{eq:At}. This provides an avenue for performing indirect 3D velocimetry in opaque, crowded droplets, where conventional PIV techniques fail or become highly cumbersome and expensive. To study the effect of confinement dimensionality, we also performed experiments of \textit{Artemia} swimming in circular (radius $R\approx0.3-2.4$ mm) quasi-2D confinement (Fig.~\ref{fig:fig4}A; hereafter denoted 2D) between two glass slides with spacing $0.48\pm0.15$ mm, chosen to restrict the swimming to 2D while allowing normal swimming motion without contact with the glass. We assume that the swimmers are constrained to swimming in a plane with a thickness corresponding roughly to their own body thickness $H$ and that they do not overlap in the $z$-direction. The dimensionless volume density is then calculated as $N^*=NV_{\mathrm{a}}/\pi HR^2\sim NWL/6R^2$, which simplifies to a dimensionless swimmer area fraction under our assumption of in-plane swimming. $N^*$ is thus a common measure of the fractional swimmer occupancy, expressed in the dimensionality accessible to the swimmers. In 3D, this quantity is naturally the swimmer volume fraction, whereas in the quasi-2D experiments, the corresponding dimensional reduction of the same occupancy concept is the projected area fraction. This formulation, however, introduces a geometric factor associated with the finite swimmer shape and the confinement thickness. The two definitions are therefore not numerically identical, but they are dimensional counterparts of the same physical measure of crowding. In the 2D experimental system, we can easily track the swimmer velocities (Fig.~\ref{fig:fig4}B and Supplementary Movie 16) and compare to the 3D results achieved with the oscillating living droplets and the scaling-law based model of Eq.~\eqref{eq:At}. We can also carefully quantify the collision kinematics between a swimmer and the 2D droplet wall (Fig.~\ref{fig:fig4}C, Supplementary Movie S17) as well as with another swimmer (Fig.~\ref{fig:fig4}D, Supplementary Movie S18).

\begin{figure*}[!t]
\center{
\includegraphics[width=0.9\linewidth]{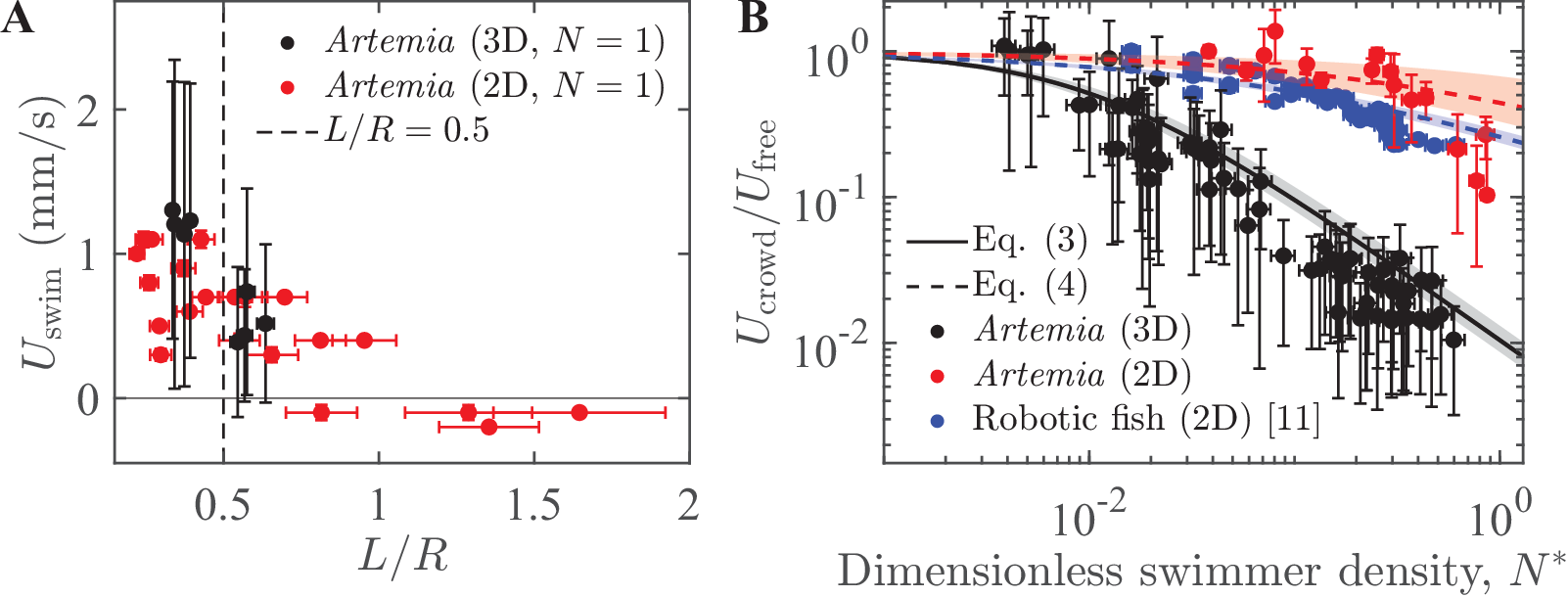}}
\caption
{\textbf{Effects of confinement and crowding in 2D and 3D.}
(A) Swimming velocity of  $N=1$ \textit{Artemia} (24h) as a function of confinement in 3D and 2D. The velocity remains unaffected at $L/R\lesssim0.5$ (dashed line). (B) Change in swimming velocity as a function of dimensionless swimmer density of \textit{Artemia} ($L/R<0.5$ and 24h) in 3D (black) and 2D (red) as well as of robotic fish in 2D \cite{Boudet2025} (blue). The solid (black) and dashed (red and blue) lines show the fits of Eq.~\eqref{eq:3} and ~\eqref{eq:4} to the 3D and 2D data, respectively. The shaded regions are the 95 \% confidence intervals of the fits. The error for $U_\mathrm{free}$ is the temporal standard deviation of several swimming cycles. The errors for the other parameters are the propagated errors with the standard deviations of their respective variables. }
\label{fig:fig5}
\end{figure*}

Since crowding and confinement may affect the swimming speed differently (as indicated by Fig.~\ref{fig:fig2}E-F), we investigate their effects separately. For uncrowded drops ($N=1$), we find that the swimming speed decreases with increased confinement (Fig.~\ref{fig:fig5}A). This is expected, as a swimmer would stop and turn when encountering an obstacle~\cite{Carkoglu2015}. From the 2D experiments, we find that an \textit{Artemia}-droplet collision takes $\tau_{\mathrm{d}} = 1.2 \pm 1.0$ s (averaged over five collision events; see example in Fig.~\ref{fig:fig4}C and Supplementary Movie S17). As a comparison, one swimming cycle lasts $0.1-0.15$ s. This behaviour will reduce the time-averaged swimming speed in confined environments. There is no clear difference between the 3D and 2D results: for both cases the speed is unaffected by the confinement at $L/R<0.5$, as assumed for the ``free'' swimmers in 3D in Fig.~\ref{fig:fig6}. At very high confinement, the swimmer cannot move normally and starts bouncing back and forth between the drop walls, rendering a net negative body velocity. This anomalous effect aligns with the strong decrease in $f_1$ with increasing $L/R$ (Fig.~\ref{fig:fig2}F).

For unconfined ($L/R<0.5$) swimmers in 2D and 3D, the swimming speed decreases with increased crowding (Fig.~\ref{fig:fig5}B). A much more pronounced decrease is observed in 3D. To rationalise this observation, we develop a minimal phenomenological model, where crowding reduces the effective swimming speed by interrupting free runs. A swimmer is assumed to move at its free-swimming speed $U_{\mathrm{free}}$ over a characteristic free-run length $l$, after which it loses a characteristic time $\tau$ because of a crowding-induced encounter. Neglecting any displacement during this interruption, the resulting effective speed $U_{\mathrm{crowd}}$ is the ratio of the typical distance travelled per cycle to the total time per run–interruption cycle 
\begin{equation}
U_{\mathrm{crowd}}=\frac{l}{l/U_{\mathrm{free}}+\tau}=\frac{U_{\mathrm{free}}}{1+\tau U_{\mathrm{free}}/l}. 
\label{eq:2}
\end{equation}
Defining the interruption rate $\lambda= U_\mathrm{free}/l$, we recast Eq.~\eqref{eq:2} as $U_{\mathrm{crowd}}/U_{\mathrm{free}}=(1+\tau\lambda)^{-1}$. Thus, the effective speed $U_{\mathrm{crowd}}$ is controlled entirely by how the encounter rate, or equivalently the characteristic run length, depends on swimmer density.

In 3D, the simplest assumption is that crowding is controlled by pairwise encounters between swimmers, with an effective rate being proportional to the swimmer number density $\lambda_{\mathrm{3D}}=k_{\mathrm{3D}}{N}/{V}$, where $k_{\mathrm{3D}}$ is an effective encounter-rate coefficient that incorporates the finite geometry of the droplet, swimmer trajectories and correlations, and the dynamical response to an encounter. We characterise crowding in the 3D droplets by the dimensionless swimmer volume fraction $N^*=NV_\mathrm{a}/V$, from which follows that $N/V=N^*/V_\mathrm{a}$, and Eq.~\eqref{eq:2} becomes 
\begin{equation}
U_{\mathrm{crowd}}^{\mathrm{3D}}(N^*)=\frac{U_{\mathrm{free}}}{1+aN^*}, \quad a = \frac{\tau_\mathrm{3D} k_\mathrm{3D}}{V_{\mathrm{a}}},
\label{eq:3}
\end{equation}
with $\tau_\mathrm{3D}$ being the characteristic interruption time. The parameter $a$ therefore combines the encounter-rate coefficient and interruption timescale, which cannot be determined separately from the present 3D measurements. At high density, when $aN* \gg1$, Eq.~\eqref{eq:3} predicts an asymptotic scaling of $U_{\mathrm{crowd}}^{\mathrm{3D}}\sim (N^*)^{-1}.$

In quasi-2D, the droplet thickness $h$ is comparable to the swimmer height $H$, and the swimmers are restricted to in-plane motion. We can therefore assume the crowding to be controlled by the in-plane spacing between swimmers. 
As explained above, the volume density $N^*$ becomes a dimensionless swimmer area fraction under this assumption, but is based on the same definition of $N^*$ as in the 3D case. 
The corresponding area number density is $\nu={N}/{A_\mathrm{c}}={N^*}/{A_{\mathrm{a}}}$, where $A_{\mathrm{a}}=\pi L W/4$ is the projected swimmer area and $A_\mathrm{c}=\pi R^2$ is the available area of the droplet. The characteristic in-plane spacing between swimmers then scales as
$\nu^{-1/2}$. Identifying the free-run length with this spacing up to an order-one geometric factor therefore gives
$l_{\mathrm{2D}}\sim \nu^{-1/2}=\sqrt{{A_{\mathrm{a}}}/{N^*}}.$ Substituting this relation into Eq.~\eqref{eq:2} gives 
\begin{equation}
U_{\mathrm{crowd}}^{\mathrm{2D}}(N^*)=\frac{U_{\mathrm{free}}}{1+b\sqrt{N^*}},\quad b\sim\frac{\tau_{\mathrm{2D}} U_{\mathrm{free}}}{\sqrt{A_{\mathrm{a}}}},
\label{eq:4}
\end{equation}
with $\tau_{\mathrm{2D}}$ being the characteristic interruption time associated with swimmer-swimmer encounters in quasi-2D. 
This equation predicts an asymptotic scaling of $U_{\mathrm{crowd}}^{\mathrm{2D}}\sim (N^*)^{-1/2}$ at high density. 

We fit Eqs.~\eqref{eq:3} and \eqref{eq:4} separately to the 3D and quasi-2D experimental \textit{Artemia} data, respectively, using one fitting parameter for each dataset. This gives $a=95\pm19$ and $b=1.3\pm0.8$ (error: 95\% confidence interval of fits). Both crossover forms describe the experimental trends well and predict different strong-crowding scalings in the two geometries. The fitted 3D coefficient also directly determines the dimensionless encounter-interruption parameter, $\tau_{\mathrm{3D}}\lambda_{\mathrm{3D}}=aN^*$. Thus, the crossover from weak to appreciable crowding occurs when $aN_{\mathrm{c}}^*\sim 1$, corresponding to $N_{\mathrm{c}}^*\sim 1/a\approx 0.01$, in agreement with our experimental observations of when the effects of crowding set in (Fig.~\ref{fig:fig6}). In 2D, an \textit{Artemia}-\textit{Artemia} collision takes $\tau_{\mathrm{a}}=0.5\pm0.2$ s (averaged over five collision events; see example in Fig.~\ref{fig:fig4}D and Supplementary Movie S18), which we take as an estimate of the interruption time $\tau_\mathrm{2D}$. Using the measured free swimming speed and swimmer dimensions, Eq.~\eqref{eq:4} gives $b\sim{\tau_{\mathrm{a}}U_{\mathrm{free}}}/{\sqrt{A_{\mathrm{a}}}}\approx 1.1$, consistent with the fitted value in Fig.~\ref{fig:fig5}B.

Finally, we test the quasi-2D model on experiments with crowded robotic toy fish by Boudet \textit{et al}.~\cite{Boudet2025}. Their data are plotted in blue in Fig.~\ref{fig:fig5}B as a function of the dimensionless area fraction $NA_{\mathrm{fish}}/\pi R^2$, where $A_{\mathrm{fish}}=\pi/4\cdot 7.6 \cdot 1.9$ cm$^2$ is the cross-sectional area of the robot swimming in a circular confinement of radius $R=15$ cm. The quasi-2D model of Eq.~\eqref{eq:4} fits the data well, with $b_{\mathrm{fish}}=2.9\pm 0.4$. Using the reported free-swimming speed of the robotic fish and assuming the geometric prefactor to be of order unity, this corresponds to an effective interruption time of $\tau_{\mathrm{fish}}\sim b_{\mathrm{fish}}\sqrt{A_{\mathrm{fish}}}/U_{\mathrm{free,fish}}\approx 1$ s. The fact that the same spacing-controlled functional form describes both \textit{Artemia} and robotic fish suggests that the underlying geometric crowding mechanism is robust across very different quasi-2D swimmer systems and Reynolds numbers (\textit{Artemia} $\mathrm{Re}=10^0-10^1$~\cite{backholm_artemia,Williams1994}, toy fish $\mathrm{Re}=10^2-10^3$~\cite{Boudet2025}). The fitted coefficient $b$, however, retains information about the system-specific swimming and encounter dynamics.

At the highest crowding levels, the measured swimming speeds fall below the corresponding theoretical predictions, potentially indicating the onset of additional many-body effects such as caging or jamming. Overall, our findings point to new opportunities for fluid control and transport at small scales using living organisms, which may enable responsive systems for sensing and actuation. Future studies that explore internal flows, collision dynamics, and active jamming could further expand the usefulness of these living droplet platforms.

\section*{Methods}\label{sec11}

\subsection*{Experimental organisms}
\label{subsec:Experimental organisms}

Both \textit{Artemia} and copepods have been used in this work (see Supplementary Fig. S1). \textit{Artemia} sp. cysts are obtained from a commercial source (Reed Mariculture Inc. CA, USA). They come in a dehydrated state and are preserved at 4~\textdegree C. The standardized protocols from Food and Agriculture Organization of the United Nations (FAO) and the International \textit{Artemia} Aquaculture Consortium (IAAC)~\cite{Stappen2024} for hatching and rearing were used throughout our experiments. \textit{Acartia tonsa} was acquired from Algova UG, Germany and is a widely distributed species across estuarine and marine environments. Adult individuals display a beat–sink swimming style and range in body length from approximately 0.75 to 1.2 mm, propulsion is carried out by means of a pair of elongated antennae in conjunction with a set of ventral thoracopods. The use of both \textit{Artemia} and copepods in research does not require ethical authorization under the applicable animal experimentation legislation of Directive 2010/63/EU. Nevertheless, animal welfare was considered throughout all experimental procedures.

Both specimens are cultured in-house at Aalto University (Espoo, Finland). The protocol for hatching is similar for both species, the only difference being the salinity of the medium. While \textit{Artemia} require salinity to be at around 40 ppt, copepods need a lower salinity between 25–30 ppt. For \textit{Artemia}, the dried cysts, stored at 4~\textdegree C, are added into a conical hatchery containing the brine medium. The medium is kept under the required conditions such as consistent bubbling, illuminated with a constant light source and temperature at 20–24~\textdegree C for obtaining a successful hatching process. \textit{Artemia} requires around 24 hours to hatch. For copepods, all the steps taken for hatching are similar with the exception of washing the cysts through a 50 µm mesh to remove the clay before adding them into to a hatchery.

Cultured copepods and \textit{Artemia} remained in the rearing tanks until transferred to the experimental setup. Experimental trials were conducted in a temperature-controlled environment at 23~\textdegree C with salinity maintained at 30 ppt for copepods and 40 ppt for Artemia, for both rearing and droplet experimental conditions. Recordings were conducted at low-light conditions, sufficient to allow for imaging and minimizing light-induced behavioural responses. The specimens are provided with a diet comprising equal proportions of the cryptophyte \textit{Rhodomonas salina} and the haptophyte \textit{Isochrysis galbana}. They are fed once every day.

At the end of each experimental trial, individuals were euthanised by immersion in $2\%$ bleach saline medium for a minimum of 24 h. Following inactivation, the material was disposed of according to the liquid-waste management procedures of Aalto University.

\subsection*{Fabrication of the black silicon superhydrophobic substrate}
\label{subsec:Fabrication of the black silicon superhydrophobic substrate}
Black silicon nanopillars were fabricated with a maskless cryogenic deep reactive ion etching process~\cite{Sainiemi2011} using an Oxford Plasmalab System 100, Oxford Instruments plasma etcher with an etching temperature of -110~\textdegree C, pressure of 10~mTorr, and etching time of 7~min. The SF$_6$ gas flow was 40~sccm, the O$_2$ gas flow 18~sccm, the ICP power was 1000~W and the forward power 6~W. The sample was made superhydrophobic by applying a plasma-enhanced chemical vapor deposited (PECVD) fluoropolymer coating (Oxford Plasmalab 80+, Oxford Instruments) at 50~W power, 250~mTorr pressure, 100~sccm of CHF$_3$, and a deposition time of 5~min.

\subsection*{Micropipette preparation}
\label{subsec:Micropipette preparation}
The protocol for manufacturing and calibrating the micropipettes used in the experiments is provided by Backholm and Bäumchen~\cite{Backholm2019}. A hollow borosilicate glass capillary tube (World Precision Instruments, USA, model TW100-6) with an outer/inner diameter of 1/0.75~mm is heated locally to soften the glass and then pulled apart creating a hollow flexible cylindrical tube of $30-50$ $\mu$m diameter. The micropipette puller used is Narishige PN-31 (Narishige International Limited, Japan) with a set parameter of main magnet 52, sub magnet 25, heater 90--95, and setscrew 5--8~mm. Changes in parameters affect the characteristics and length of micropipettes and hence are kept consistent. The spring constant of these micropipettes is highly dependent on the length of the cantilever. The length is adjusted by cutting the tip using the MF2 microforge (Narishige 11 International Limited, Japan). In this process, the micropipette puller is brought in contact with a uniformly heated small spherical glass bead allowing the softened micropipette to adhere to it. The temperature is cut off, cooling the glass bead, and the sudden temperature gradient causes the micropipette to break slightly above the point of contact, obtaining a smooth rim. Micropipettes with the lengths of $2.5-2.8$ cm are used in all experiments of this work. 

\subsection*{Calibration of the micropipettes}
\label{subsec:Calibration of the micropipettes}
The micropipettes are calibrated (see Supplementary Fig. S5) to obtain the spring constant using the water droplet method~\cite{Backholm2019}. Water is injected through a horizontally mounted micropipette till a small water droplet is formed on the tip, the weight of which causes the micropipette to deflect downward. The deflection is recorded using a camera (FLIR Grasshopper3 GS3-U3-23S6M-C, Teledyne Vision Solutions, USA) at 40--100~fps, then analysed in MATLAB. Assuming that the water droplet is an ellipsoid, the volume $V = {\pi l_\text{min}^2 l_\text{max}}/{6}$, where $l_\text{min}$ and $l_\text{max}$ are the minor and major axes of the ellipsoidal droplet, is calculated through image analysis. By comparing the weight force of the droplet with the corresponding deflection, we obtain the spring constant, $k = {\rho V g}/{x}$ $(\rho = 1000~\text{kg/m}^3)$.

\subsection*{Experimental setup and data acquisition}
\label{subsec:Experimental setup and data acquisition}
The experimental setup for the 3D experiments (see Supplementary Fig. S7) consists of a vibration-isolated table (Halcyonics i4large, Accurion) completely shielded by black hardboard box (Thorlabs) to prevent airflow which strongly affects the data acquired. The superhydrophobic substrate is positioned on a modular stage comprising of $xyz$-translational stage (Thorlabs) for precise spatial adjustments with a goniometer (Thorlabs) mounted on top to achieve perfect horizontal surface. The perfectly horizontal surface ensures that the droplet remains on the substrate without rolling off. The droplet is imaged using a camera (FLIR GS3-U3-130 23S6M-C, Integrated Imaging Solutions, Inc.) connected to an objective (2x, Mitutoyo 378-801-12, Edmund Optics, UK) through the horizontal video microscopy unit, VMU (Mitutoyo 378-506, VIS-NIR range, Edmund Optics, UK), with the droplet being illuminated from the back using an LED (BT100100-WHIIC, Edmunds Optics, UK) light source. The micropipette is positioned symmetrically just inside the droplet interface from the top for consistent lateral deflection measurements as shown in Fig.~\ref{fig:fig1}. The deflection of the micropipette is recorded in the form of image sequences obtained from the camera, recorded at a sampling rate of \(f_s=40~\mathrm{Hz}\) for a recording time window of \(T=60~\mathrm{s}\), giving \(N=2400\) data points. The Nyquist frequency was therefore \(f_s/2=20~\mathrm{Hz}\), which is higher than the measured droplet oscillation frequencies. The frequency resolution was \(\Delta f=1/T\approx0.017~\mathrm{Hz}\). After the living droplet experiments, the drop is moved to a glass slide, which is placed on an inverted microscope where the swimmer sizes and numbers are determined.

The experimental setup for the 2D experiments is built using optomechanical parts (Thorlabs) on a passively damped optical table (ST Instruments, Netherlands). The sample stage is mounted on a dampend post and an optical post assembly is mounted to the sample stage to hold a light source (SL244, Advanced Illumination, USA). The light source is covered by a precision wipe (05511, Kimtech Science, USA) to act as a diffuser. The high-speed camera (PhantomMiro C211, AMETEK, Inc., USA) is mounted below the sample stage with a video microscopy unit (Mitutoyo) and a 1x Plan Apo Infinity Corrected Long WD objective (Mitutoyo, Japan). The camera is connected to the ethernet port of the computer running the Phantom PCC 3.8 software. On another damped post, the micropipette holder and syringe system (Narishige International Limited, Japan) is operated with micromanipulators (MN-153, Narishige International Limited, Japan). This pipette system consists of a 10 mL luer-lock plastic syringe, half filled with brine water, attached to an HI-7 injection holder (Narishige International Limited, Japan) connected with PTFE plastic tubing, ending in a stainless-steel pipette holder. A glass micropipette is inserted into this holder. The micropipette system is mounted in parallel with the sample stage. The sample is prepared by placing a 75 mm x 50 mm glass microscope slide (2947-75X50, Corning, USA) on top of a sample holder. Four pieces cut from a 0.3 mm copper sheet are used as spacers. Using a Finnpipette F2 (Finn 4662000, Thermo Scientific, USA), a drop with swimming \textit{Artemia} is placed on the glass slide between the four spacers. Finally, a 24 mm x 50 mm flexible cover slip (VWRI631-0147, Avantor, USA) is placed over the spacers and in contact with the droplet, forming a capillary bridge. The micropipette is slid between the two glass slides and is used to adjust the droplet volume in a minimally invasive way. Experiments are recorded at 100 fps.

\subsection*{Declaration of AI use in figure preparation}
The schematic images in Fig.~\ref{fig:fig1}A and Fig.~\ref{fig:fig4}A were created with assistance from OpenAI’s ChatGPT (GPT-5). We have reviewed and verified the schematics to ensure scientific accuracy and take full responsibility for the final content.

\backmatter

\bmhead{Supplementary information}
The online version contains supplementary material available at
https://doi.org/xx.xx/xxxxxx-xxx-xxxxx-x.

\bmhead{Funding}
This work was funded by the European Union (ERC StG SWARM, 101115076; M.B.), the Research Council of Finland (MesoSwim, 354904; M.B.), and the National Science Centre of Poland (Sonata Bis grant no. 2023/50/E/ST3/00465; M.L.). Views and opinions expressed are however those of the authors only and do not necessarily reflect those of the European Union or the European Research Council. Neither the European Union nor the granting authority can be held responsible for them.

\bmhead{Competing interests}
The authors declare no competing interests.

\bmhead{Data availability}
The datasets used to plot graphs in the paper and examples of raw data files are shared on Zenodo~\cite{MalikZen25}.

\bmhead{Author contribution}
M.B. conceived and supervised the project and acquired the funding. L.M., S.N., C.P.L., M. Lamminmäki, and R.A.L. did experiments. L.M., S.N., M. Lamminmäki, and R.A.L. fabricated the micropipettes. R.A.L. helped in hatching living organisms. M.B., S.N., L.M. and C.P.L. wrote and/or modified the MATLAB codes for data analysis. L.M., S.N., and M.B. analysed the data. L.M. and M.B. did plotting and made figures. L.M. and C.P.L. did video formatting. V.J. provided superhydrophobic substrates and gave suggestions. M.B. and M.L. did theoretical modelling. L.M., M.L., and M.B. wrote the paper with input from all authors.

\noindent

\bibliography{sn-bibliography}

\end{document}